\documentclass[%
 aip,
 amsmath,amssymb,
 reprint,%
]{revtex4-1}

\usepackage{graphicx}
\usepackage{dcolumn}
\usepackage{bm}

\usepackage[utf8]{inputenc}
\usepackage[T1]{fontenc}
\usepackage{mathptmx}
\usepackage{etoolbox}
\usepackage{physics}

\makeatletter
\def\@email#1#2{%
 \endgroup
 \patchcmd{\titleblock@produce}
  {\frontmatter@RRAPformat}
  {\frontmatter@RRAPformat{\produce@RRAP{*#1\href{mailto:#2}{#2}}}\frontmatter@RRAPformat}
  {}{}
}%
\makeatother
\begin{document}

\preprint{AIP/123-QED}

\title{Full-band Monte Carlo simulation of two-dimensional electron gas in (Al$_{x}$Ga$_{1-x}$)$_{2}$O$_{3}$/Ga$_{2}$O$_{3}$ heterostructures}
\author{Avinash Kumar}
\email{a42@buffalo.edu.}
 \affiliation{Department of Electrical Engineering, University at Buffalo, The State University of New York.}
\author{Uttam Singisetti}%
 \email{uttamsin@buffalo.edu.}
  \affiliation{Department of Electrical Engineering, University at Buffalo, The State University of New York.}


\date{\today}

\begin{abstract}
$\beta$-Gallium Oxide (Ga$_{2}$O$_{3}$) is an extensively investigated ultrawide-bandgap semiconductor for potential applications in power electronics and RF switching. The room temperature bulk electron mobility ($\sim$200 cm$^{2}$V$^{-1}$s$^{-1}$) is comparatively low and is limited by the 30 phonon modes originating from its 10-atom primitive cell. The theoretically calculated saturation velocity is 1-2$\times$10$^{7}$ cms$^{-1}$ which is comparable to GaN. The high field electron transport in the 2DEG is explored in this work based on the first principles calculated parameters. A self-consistent calculation on a given heterostructure design gives the confined eigenfunctions and eigenenergies. The intrasubband and the intersubband scattering rates are calculated based on the Fermi's golden rule considering LO phonon-plasmon screening. The high field characteristics are extracted from the full-band Monte Carlo simulation of heterostructures at 300 K. The motion of electrons in the 2DEG and the bulk is treated through an integrated Monte Carlo program which outputs the steady state zone population, transient dynamics and the velocity-field curves for a few heterostructure designs. The critical field for saturation does not change significantly  from bulk values, however an improved peak velocity is calculated at a higher 2DEG density. The velocity at low 2DEG densities is impacted by the antiscreening of LO phonons which plays an important role in shaping the zone population. A comparison with the experimental measurements is also carried out and possible origins of the discrepancies with experiments is discussed. 
\end{abstract}

\maketitle

\section{\label{sec:level1}Introduction}

$\beta-$Ga$_{2}$O$_{3}$ is a promising wide-bandgap semiconductor material known for its potential applications in high voltage power electronics\cite{ref96,ref97,ref98,ref99,ref100,ref101,ref102,ref103,ref104,ref105} and high power radio frequency (RF) switching\cite{ref106,ref107,ref108,ref109}. The power electronics application comes from its large Baligas's figure of merit (BFoM = $\epsilon\mu E_{c}^{3}$) resulting from its wide bandgap (4.8 eV)\cite{ref86,ref87,ref88,ref89} and a very high estimated breakdown field ($E_{c}$ = 8 MVcm$^{-1}$)\cite{ref26}. The RF performance of a material is characterized by its Johnson's figure of merit, given by JFoM = $ E_{c}v_{s}/2\pi$, where $v_{s}$ is the saturation velocity, which in turn is limited by the low field electron mobility. The bulk mobility in the corresponding devices is low, due to dominant multiple polar optical modes present at 300 K\cite{ref4,ref125,ref126,ref127,ref128}, which is shown to be improved in the 2DEG of AlGa$_{2}$O$_{3}$/Ga$_{2}$O$_{3}$ heterostructures in our previous work\cite{ref15}. There are several experimental measurements on modulation doped transistor of AlGa$_{2}$O$_{3}$/Ga$_{2}$O$_{3}$ reporting low field electronic mobility at low and room temperature\cite{ref16,ref17,ref27,ref137,ref138,ref165}. Zhang et al.\cite{ref17} recently reported velocity-field characteristics of such heterostructures at 50 K and the corresponding low field mobility and saturation velocity are found to be $\sim1500$ cm$^2$V$^{-1}$s$^{-1}$ and 1.1$\times$10$^{7}$ cms$^{-1}$ respectively. The high field electron transport in a bulk material has been reported previously\cite{ref18}. Yan Liu et al.\cite{ref200} recently reported Monte Carlo based velocity-field characteristics of unintentionally doped heterostructures based on analytically calculated scattering rates. However, there are no reports for the 2DEG which takes into account the first principles calculated full band electron-phonon interaction elements. It is of high interest to incorporate those parameters in order to completely understand the behaviour of heterostructures when subjected to high field to fully take the advantage of such devices. 

A complete \textit{ab-initio} study of $\beta-$Ga$_{2}$O$_{3}$ is very challenging as compared to other wide-bandgap materials like GaN due to it's low crystal symmetry and large primitive cell size. However, a full band Monte Carlo (FBMC) investigation is required with the inclusion of true electron-phonon interaction (EPI) elements to capture the band anisotropy as well as non-parabolic effects. 

We discuss a basic flow of methods used for our Monte Carlo simulation in the next section. The calculation starts with first principles calculation on electronic band structure and phonon dispersion followed by short range and long range electron phonon interaction element calculation. This is followed by finding a self-consistent solution of the device in consideration which outputs necessary quantum well parameters required for the scattering rate calculation. We then discuss the methods and equations used for 2D $\&$ 3D scattering rate calculation followed by a discussion on different scattering rates variation.

In the third section, we discuss the FBMC simulation along with different steps involved in the process. We discuss the equilibrium distribution of an ensemble of electrons when no field is applied. Next, we discuss the normalized scattering rate needed to select a random scattering process during the simulation. Finally, in this section, we discuss the final state election methods and their implementation for different scattering mechanisms and transport regime.

The results extracted from FBMC simulation are discussed in the third section. We discuss the zone population at different fields in energy and k-space. This is followed by a discussion on transient dynamics and velocity field characteristics for a few cases. We also present a comparison with an experimental device and discuss the possible reasons of discrepancies observed.

We finally conclude our work in the fifth section with a quick look on important results and discuss any suggestions for improvements in current heterostructure devices to help the experimental community.

\section{\label{sec:level2}Methodology}

\begin{figure}
\includegraphics[width=0.45\textwidth]{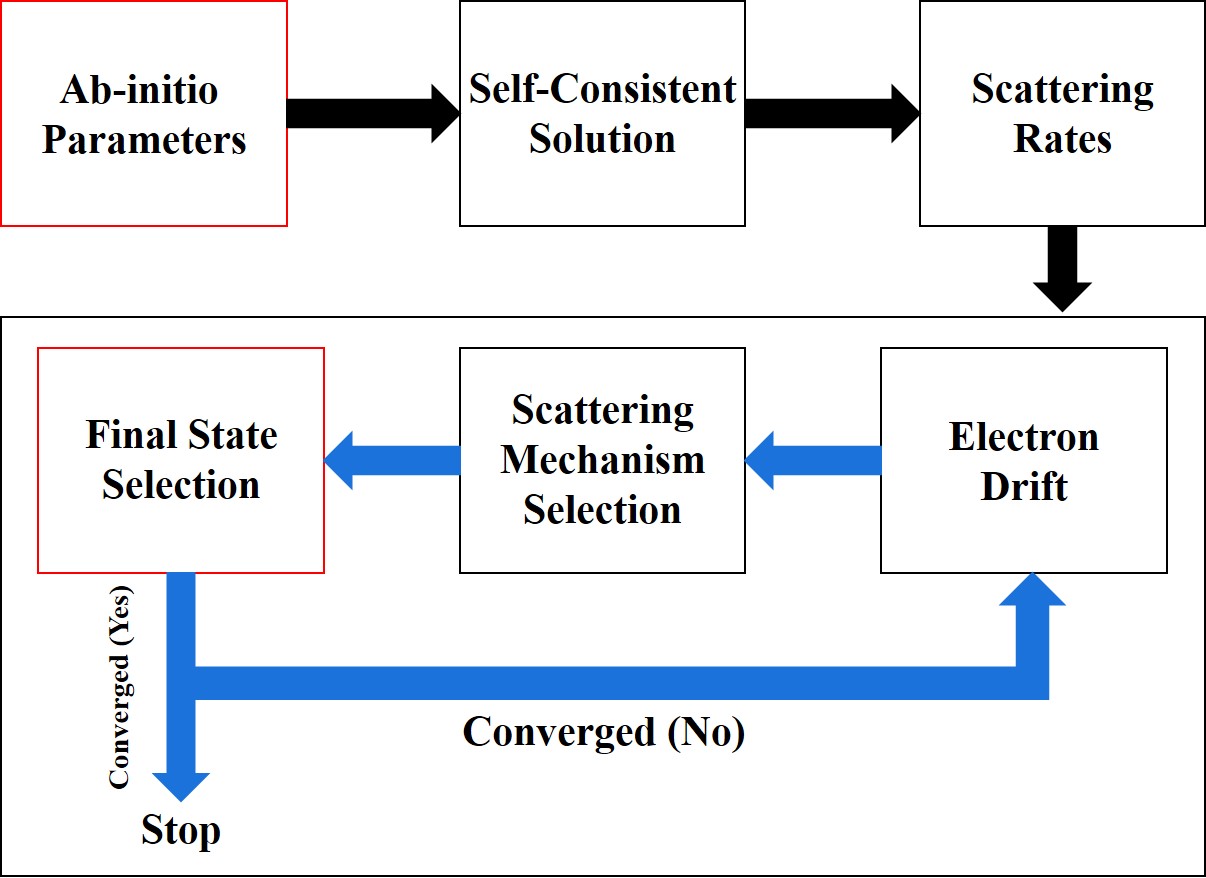}
\caption{\label{fig1} A flow chart showing the different steps involved in the Monte Carlo simulation. The boxed (red) steps are the most computationally expensive and require high performance computing resources. Each step is treated separately and finally combined to work in a combination of series and parallel simulation.}
\end{figure}

Fig.\ref{fig1} shows a basic flow chart of the calculations involved in the high field electron transport simulation. The process starts with calculating the \textit{ab-initio} parameters from well established density function theory (DFT) and density functional perturbation theory (DFPT) calculations using existing tools and techniques. This gives the electronic band structure, phonon dispersion and electron-phonon interaction elements used later scattering rate calculations and in FBMC simulation. This is followed by finding self-consistent solution of the heterostructure in consideration from which the 2DEG and the device parameters are extracted. The earlier two steps provides us with the necessary parameters for the scattering rate calculation which is the next step in the process. The last step is to run the FBMC simulation until a convergence in an observable is achieved. The FBMC involves three main subprocesses: electron drift, randomly selecting a scattering even and the corresponding final state selection. We next provide a brief description of first principles, self-consistent and the scattering rate calculation. The FBMC is, however, discussed in the next section to provide a better understanding of the method given its computational complexity.

\subsection{\label{sec:level2.1}First principles calculations}

\begin{figure}
\includegraphics[width=0.5\textwidth]{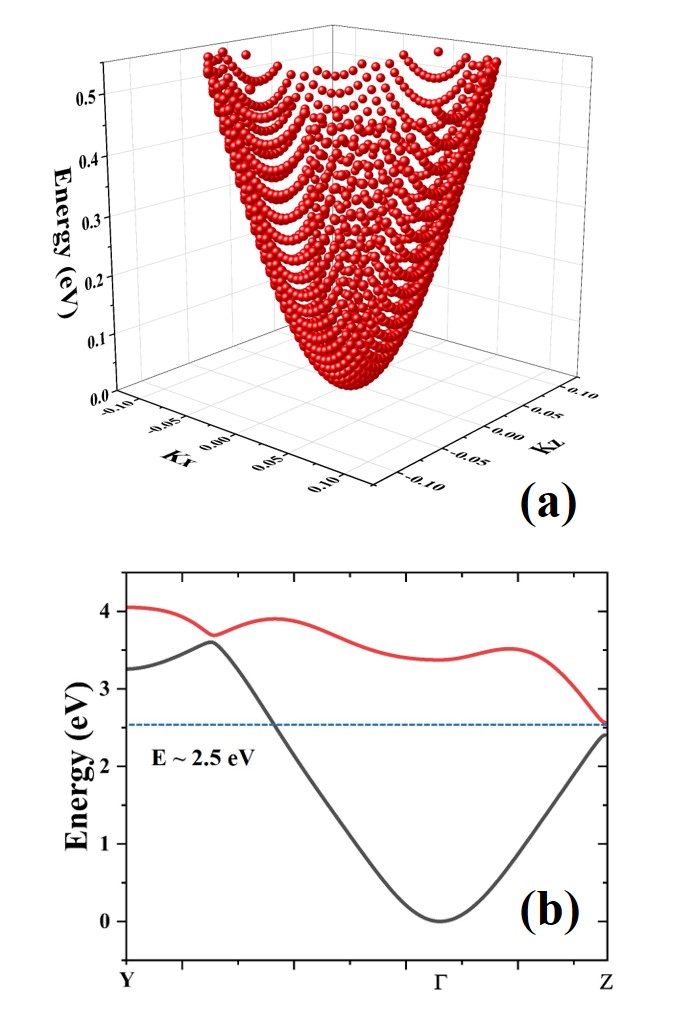}
\caption{\label{fig2} (a) The DFT calculated electronic band structure (first conduction band upto 0.5 eV) for $\mathrm{\beta-Ga_{2}O_{3}}$ on a 2D k-mesh of grid size 200$\times$200 in the entire brillouin zone. Here, $k_{x}$ and $k_{y}$ represent cartesian directions. The bottom of the conduction band is assumed to be at 0 eV. The isotropy in the band can be clearly seen. (b) The first two conduction bands in two specific directions Y(0,0.5,0) and Z(0,0,0.5) from the $\Gamma$(0,0,0) point in the reciprocal space are shown. A smooth plot is obtained through Wannier interpolation on a dense grid size of 80$\times$80$\times$80 in the entire brillouin zone. The calculations are done using Quantum Espresso \cite{ref10}.}
\end{figure}

The electronic band structure, as shown in fig.\ref{fig2}, is first calculated and Wannier interpolated on a fine k-grid using Quantum Espresso\cite{ref10} followed by Wannier90 package\cite{ref12}. A 80$\times$80$\times$80 k-mesh is used for the electrons in the 3D regime and a more denser k-mesh of 200$\times$1$\times$200 is used for the electrons in the 2D regime. The 2DEG is assumed to be confined in the cartesian y-direction. The 2D-bandstructure in fig.\ref{fig2}(a) is plotted in x-z cartesian space and the 3D-bandstructure in fig.\ref{fig2}(b) is shown along Y(0,0.5,0) and Z(0,0,0.5) reciprocal space directions. Only first two conduction bands are taken into account following a previous work\cite{ref18} on a bulk system where the distribution at moderately high fields drops rapidly after $\sim$2 eV. The conduction band minimum is fairly isotropic with electron effective mass in the range (0.27-0.3)m$_{e}$. The non-parabolicity in the band structure starts to appear at higher energies. It is important to point out that the first satellite valley lies around $\sim$2.5 eV, way above the conduction band minimum. 

The long range nature of polar optical phonons (POP) is properly captured with a very dense grid of 40$\times$40$\times$40 covering just the 40$\%$ of the full brillouin zone. In order the to incorporate the dynamic screening of POP through 2D-plasmons, the electron - LO phonon interaction elements are calculated using Fr\"{o}hlich interaction following our earlier works\cite{ref15,ref184}. In addition, the matrix elements are required to be stored for each $k$, each $q$ and for every mode $\nu$ for later use. POP modes do not mediate any inter-band scattering and are $k$ independent due to their long range nature (small $q$) and hence are much easier to handle. The POP EPI elements are calculated separately as described in the next section, with first principles parameters such as LO-TO frequencies, dielectric tensor ($\epsilon_{\infty}$) and displacement vectors as input extracted from DFT and  DFPT calculations. The pure LO modes corresponding to each $q$ are calculated by diagonalizing the DFPT computed dynamical matrix at $\Gamma$ point with macroscopic polarization added. 

The short range electron-phonon interaction elements (Acoustic and Non-polar optical phonon) are calculated on a 40$\times$40$\times$40 q-grid in the entire brillouin zone using EPW package\cite{ref11}. The non-polar EPI elements are calculated using EPW package through Wannier interpolation of matrix elements on a coarse mesh. For low energies, the contribution to the non-polar matrix elements coming from the overlap integral of the periodic part of the Bloch wavefunction becomes 1 due to spherical and isotropic bands. This enables us to directly use the 2D scattering rate equation. For higher energies, when the electron is in the bulk, there are no such approximation as the corresponding matrix elements are highly anisotropic and must be considered with a proper care. For non-polar EPI elements, the initial and final state bands are also stored, exponentially increasing the memory requirement. The idea to reduce the space requirement is to only store those matrix elements where energy-momentum conservation is satisfied as the rest will be anyway rejected during the final state selection in the Monte Carlo technique (discussed later). This is done with a little modification in the EPW code. In order to minimize RAM requirements, the non-polar matrix elements are divided and stored in multiple files and read through I/O process during the simulation. This is still computationally intensive as during a parallel run, multiple cores could access the files at the same time and abruptly increase the RAM requirement. This is optimized through trading off the number of files as more number number of files would take more time to read which is again not desired. Note that total memory requirement (main memory) remains the same with number of files.

\subsection{\label{sec:level2.2}Self-consistent solution}

\begin{figure}
\includegraphics[width=0.5\textwidth]{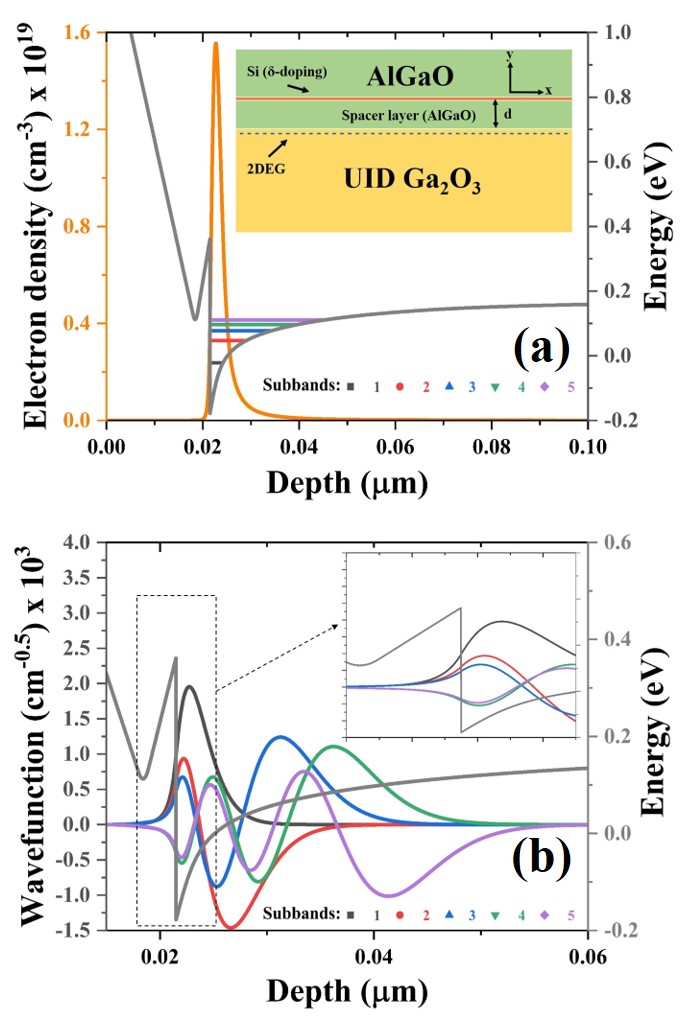}
\caption{\label{fig3} (a) The band diagram (solid gray, right-axis) showing the quantum well near the interface where the 2DEG is formed, and the orange line (left-axis) showing the variation of electron density in the heterostructure, which can be clearly seen to peak in the confined region. The inset shows the $\beta$-(A\MakeLowercase l$_{\MakeLowercase x}$G\MakeLowercase a$_{1-\MakeLowercase x}$)$_{2}$O$_{3}$/G\MakeLowercase a$_{2}$O$_{3}$ heterostructure used in this work. The channel lies in the x-z plane and the confinement is in the cartesian $y$ direction. A delta doping is assumed at a distance $d$ from the interface, where the 2DEG is formed. The bulk at the bottom is unintentionally doped with a doping density of 1$\times$10$^{16}$ cm$^{-3}$. (b) The band diagram (solid gray, right-axis) along with the confined wavefunctions (left axis) corresponding to the first 5 subbands (colored as shown). The axes are scaled to show a clear comparison. The delocalization of electrons is apparent moving from the black solid line (subband 1) to purple solid line(subband 5). The inset shows a close up near the interface. This heterostructure corresponds to an electron density of 5$\times$10$^{12}$ cm$^{-2}$, 3 nm spacer layer and 20$\%$ Al in the alloy. }
\end{figure}

\begin{table}
\caption{\label{tab1}Material parameters used in self-consistent calculation of $\beta$-(Al$_{0.2}$Ga$_{0.8}$)$_{2}$O$_{3}$/Ga$_{2}$O$_{3}$ heterostructures. The data is taken from \cite{ref28,ref137,ref30,ref31,ref32}.}
\begin{ruledtabular}
\begin{tabular}{lr}
   $m^{*}$ & 0.3$m_{e}$\\
   Bandgap $E_{g}$ ($\beta$-Ga$_{2}$O$_{3}$) & 4.7 eV\\
   Bandgap $E_{g}$ ($\beta$-(Al$_{0.2}$Ga$_{0.8}$)$_{2}$O$_{3}$) & 5.0 eV\\
   Band-offset $\Delta$E (AGO/GO) & 0.54 eV\\
   Donor energy $E_{d}$ ($\beta-$(Al$_{0.2}$Ga$_{0.8}$)$_{2}$O$_{3}$) & 0.135 eV\\
   Dielectric constant $\epsilon_{r}$ ($\beta$-Ga$_{2}$O$_{3}$) & 10\\
\end{tabular}
\end{ruledtabular}
\end{table}

A heterostructure device, as shown in the inset of fig.\ref{fig3}(a), is used as a reference with 20$\%$ aluminum concentration in the alloy and impurities doped at 3 nm (spacer thickness) away from the interface. However, the spacer thickness is adjusted to 4.5 nm to provide a fair comparison with the experiment. Table \ref{tab1} shows a list of material parameters used in our simulation. As seen in the fig.\ref{fig3}(a), a 2DEG is formed at the interface, confined in a quantum well. The delta doping is adjusted for each case to obtain a given 2DEG density in the channel. The effects of quantum confinement is modeled through the solution of Schr\"odinger equation, given by eq.(\ref{eq3.1}), along with other fundamental device equations. 

\begin{equation}\label{eq3.1}
    {-\frac{\hbar^2}{2m^*}\frac{d^2\psi_{n}(y)}{dy^2}+V(y)\psi_{n}(y)=E_{n}\psi_{n}(y)}
\end{equation}
Where, ${V(y)=-e\phi_{e}(y)+V_{h}(y)}$ is the effective potential. Here, ${V_{h}(y)}$ is the step potential barrier at the interface, and ${\phi_{e}(y)}$ is the electrostatic potential. 

This provides the quantized density of states under the influence of quantum well potential. Here, $E_{n}$ and $\psi_{n}$ are the $n^{th}$ bound state energy and wavefunction respectively. A similar set of equations can be used to describe holes. 

The discrete nature of the quantized density of states allows to reduce the integral to a sum over bound state energies and the electron density can be given by::

\begin{equation}\label{eq3.2}
    {N_{i}=\frac{m^*k_{B}T}{\pi\hbar^2}ln\Bigg[1+e^{\Big(\frac{E_{F}-E_{i}}{k_{B}T}\Big)}\Bigg]}
\end{equation}
This can be, in turn, used as an input to the Poisson equation given by:

\begin{equation}\label{eq3.3}
    {\frac{d^2\phi_{e}(y)}{dy^2}=\frac{e}{\epsilon_{0}\epsilon_{r}}\Bigg[\sum_{i} N_{i}\psi_{i}^{2}(y)+N_{A}(y)-N_{D}(y)\Bigg]}
\end{equation}
Here, $e$ is the electronic charge, ${\epsilon_{o}}$ is the permittivity of free space and ${\epsilon_{r}}$ is the dielectric constant. ${N_{i}}$ is the number of electrons in the $i^{th}$ subband. ${N_{A}(y)}$ and ${N_{D}(y)}$ are the acceptor and donor concentrations respectively. Also, ${E_{F}}$ is the Fermi energy, ${k_{B}}$ is the Boltzmann constant and $T$ is the temperature.

The solution to this equation provides potential which can be further substituted into Schr\"odinger equation for the next iteration. This keeps going until a convergence is achieved and a self-consistent solution of Schr\"odinger-Poisson equation is found.

The above equations are solved using Silvaco Atlas\cite{ref14} for the heterostructure used in this work and the eigenvalues and eigenvectors corresponding to the first five subbands are found. The electrons with higher energy (above fifth subband) are no more confined and have finite probability density away from the interface, as shown in fig.\ref{fig3}(b), and hence assumed to be bulk like. The inset of fig.\ref{fig3}(b) shows a close-up of the wavefunctions near the interface. As we move higher in the bands, the electrons becomes more bulk-like and start moving away from the interface into the bulk as they are no more confined. This reduces the corresponding probability density near the interface which is expected to affect the scattering rates as we will see in the next section. The transverse electric field at the interface is also extracted to calculate interface roughness scattering.

\subsection{\label{sec:level2.3}Scattering rates}

\begin{figure}
\includegraphics[width=0.45\textwidth]{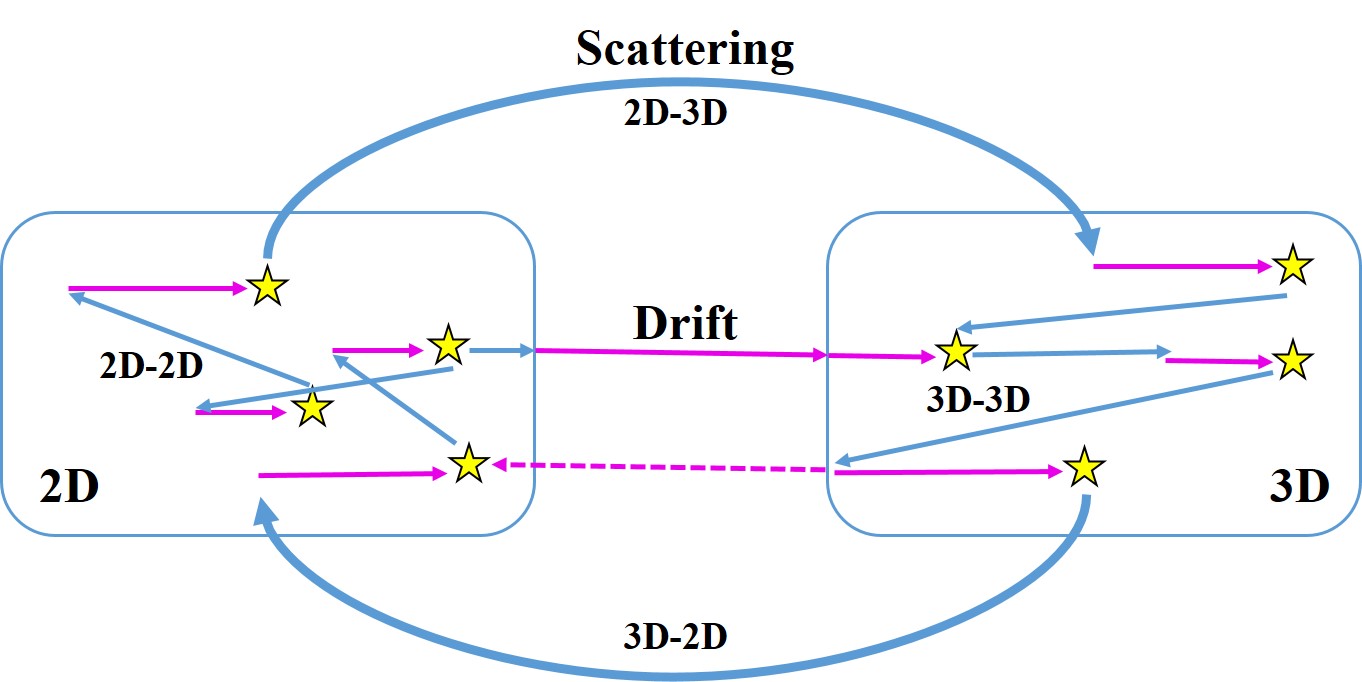}
\caption{\label{fig4} A schematic of the high field electron transport in a heterostructure. The transport happens in a 2D and a bulk region with the transitions mediated through drift and scattering of carriers represented by the magenta and blue arrows (preceded by a yellow star) respectively. The dashed magenta arrow represents a case when the 3D-2D transition is mediated through the carrier drift. This happens when a carrier in the bulk gets its momentum reversed (opposite to electric field) through a scattering process.}
\end{figure}

Electrons in the heterostructure devices are subject to scattering from remote impurities, interface roughness, irregularities in the alloy and the phonons. An accurate desciption of such scattering mechanisms is critical for high field transport study. All the major elastic scattering mechs. such as alloy disorder, remote impurity and interface roughness as well as inelastic scattering mechs. such as polar optical, non-polar optical and acoustic phonon scatterings are taken into account. 

Fig.\ref{fig4} shows a schematic of carrier transport under high electric field. The whole system is divided into a 2D and a 3D regime based on a cutoff-energy \cite{ref201}. Electrons in each region experience drift by the applied electric field and scattering by the perturbed potentials. Unlike the electron transport under low field, the high-field could push the electrons to higher subbands and even to higher conduction bands once the electron has transitioned to the bulk. Hence, we must consider the intersubband scattering processes when the electron is confined and interband scattering processes once it has escaped the confinement. The scattering in the 2DEG is limited by an energy cutoff (0.3 eV), where the 2D-3D transition happens \cite{ref202,ref203,ref204,ref205,ref206}. This energy cutoff is decided based on the confinement potential energy coming from the Schr\"odinger-Poisson solution. The 2D-3D $\&$ 3D-2D transitions are mediated through scattering as well as drift \cite{ref201}. five subbands are taken into account below 0.3 eV in the 2DEG as the higher subbands have very small energy difference with bulk characteristics. Only intra-subband transitions are screened in case of elastic processes as all inter-subband transitions would require higher $q$ where the screening factor vanishes. The POP modes are dynamically screened as described later in this section. The non-polar modes are calculated from first principles and each mode is treated separately while calculating the corresponding 2D scattering rates. Here, acoustic modes and non-polar optical phonon modes are treated together and hence we use non-polar to refer them all for convenience. This eliminates any curve fitting and hence the deformation potential constants and also prohibits the use of Thomas-Fermi type screening. To avoid any related errors, the non-polar transitions are kept unscreened. Also, since the non-polar scattering is a high q-process, the screening becomes weak and ineffective, justifying our assumption.

For the electrons in the bulk ($>$ 0.3 eV), the scattering is through POP and non-polar (non-polar optical plus acoustic) modes only as any ionized impurity scattering drops quickly at higher energies and would not contribute much at higher fields. As the 2DEG is confined in a quantum well, limited by a given energy, the average electron density (over the channel region) drops rapidly in the bulk and so the scattering processes in the same are assumed to be free of any screening. The POP EPI elements are then calculated assuming low electron density in the bulk such that the plasmon energy is way below the range of LO phonon energies with no LOPC or dynamic screening present. First 2 conduction bands are considered \cite{ref18} and the inter-band scattering is assumed to be only mediated by non-polar modes as, due to POP scattering being a small $q$ process, the overlap between the two wavefunctions (orthogonal) vanishes and the corresponding scattering rate becomes negligible.

The scattering rates are calculated based on Fermi's golden rule\cite{ref192} as given by:

\begin{widetext}
\begin{equation} \label{eq4.17}
S(\vec{k}_{i},\vec{k}_{f}) = \frac{2\pi}{\hbar} \abs{H_{\vec{k}_{f}\vec{k}_{i}}^{a}}^{2}\abs{I_{mn}}^2\delta(E(\vec{k}_{f})-E(\vec{k}_{i})-\hbar\omega)
+ \frac{2\pi}{\hbar} \abs{H_{\vec{k}_{f}\vec{k}_{i}}^{e}}^{2}\abs{I_{mn}}^2\delta(E(\vec{k}_{f})-E(\vec{k}_{i})+\hbar\omega)
\end{equation}
\end{widetext}

Where, $I_{mn}=\int_{-\infty}^{\infty}\psi_{n}(y)e^{\pm q_{y}y}\psi_{m}^{*}(y)dy$, where $n$ and $m$ correspond to initial and final subband respectively. This is a general equation with a strict momentum conservation in only 2D space and an extra term containing the overlap of initial and final state wavefunctions represents momentum conservation in the third direction \cite{ref193}. This is limited by the uncertainty principle, meaning a fuzziness in the momentum conservation if the third direction is confined (2DEG) and a strict momentum conservation otherwise (bulk). The energy conservation is still the total energy conservation.

\subsubsection{Elastic scattering rates: Remote impurity, Interface roughness $\&$ Alloy disorder (2D-2D)}

The remote impurity momentum relaxation rate is then calculated as \cite{ref169}:

\begin{equation}\label{eq4.21}
    {\frac{1}{\tau_{RI}}=\frac{m^*e^4Z^2}{8\pi\hbar^3\epsilon_{r}^2\epsilon_{0}^2}\int_{0}^{2\pi}\Big(\frac{F(q,y_{i})}{qS(q)}\Big)^2N(y_{i})(1-cos\theta)dy_i}
\end{equation}
where \textit{Ze} is the charge on ionized impurities, ${\epsilon_{r}}$ is the static dielectric constant, ${N(y_{i})}$ is the impurity distribution, and

\begin{equation}\label{eq4.22}
    {F(q,y_{i}) = \int_{-\infty}^{\infty}\psi_{n}(y)e^{-q|y_{i}-y|}\psi_{m}^{*}(y)dy}
\end{equation}

The Interface roughness momentum relaxation rate is given by \cite{ref194}:

\begin{equation}\label{eq4.23}
    {\frac{1}{\tau_{IFR}}=\frac{m^*e^2E_{eff}^2\delta^2 L^2}{2\hbar^3}\int_{0}^{2\pi}\frac{e^{-\frac{q^2L^2}{4}}(1-cos\theta)}{S(q)^2}d\theta}
\end{equation}
where,
\begin{equation}\label{eq4.24}
    {E_{eff} = \int_{-\infty}^{\infty}\psi_{n}(y)\frac{dV}{dy}\psi_{m}^{*}(y)dy}
\end{equation}
Here ${\frac{dV}{dy}}$ is the electric field along the confinement direction, which pushes the electrons to collide with the interface.

The alloy disorder momentum relaxation rate is calculated using \cite{ref196}:

\begin{equation}\label{eq4.25}
    {\frac{1}{\tau_{Alloy}}=\frac{m^*\Omega_{o}x(1-x)(\delta E_{c})^2F_{al}}{2\pi\hbar^3}\int_{0}^{2\pi}\frac{1-cos\theta}{S(q)^2}d\theta}
\end{equation}

\begin{equation}\label{eq4.26}
    {F_{al} = \int_{-\infty}^{0}|\psi_{n}(y)|^2|\psi_{m}(y)|^2dy}
\end{equation}
The unit cell volume at each Al concentration is taken from \cite{ref31} by interpolating the available data. The scattering potential is assumed to be equal to the conduction band offset $\delta E_{c}$ between $\mathrm{Al_{2}O_{3}}$ and $\mathrm{Ga_{2}O_{3}}$ \cite{ref31}. 

\begin{figure*}
\includegraphics[width=\textwidth]{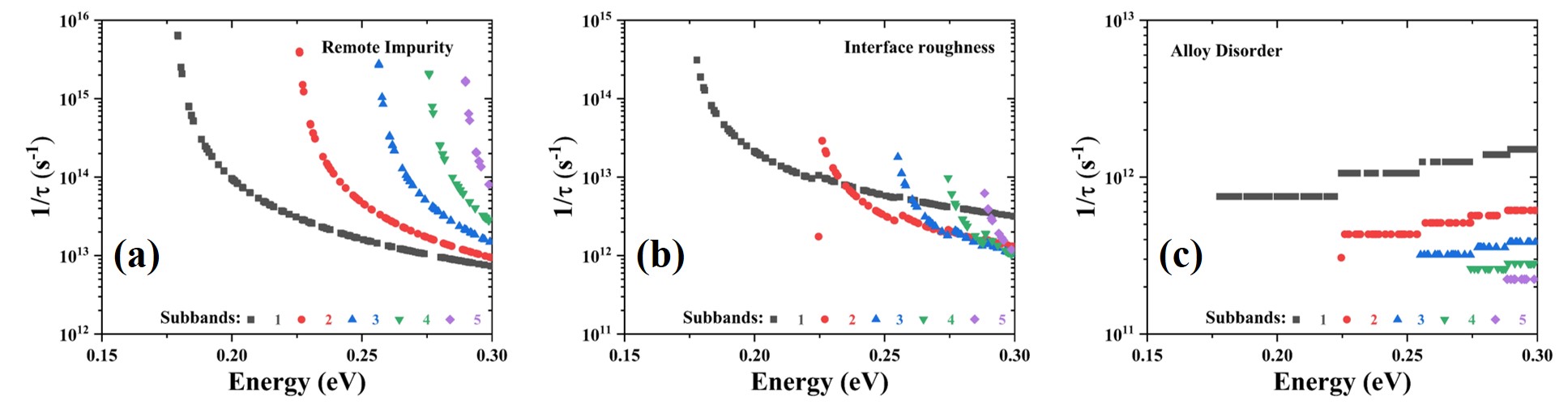}
\caption{\label{fig5} The 2D-2D (a) remote, (b) interface roughness, and (c) alloy disorder scattering rates (300 K) as a function of electron energy (eV) for the first 5 subbands. The 0 eV is assumed to be the bottom of the 1st conduction band. The scattering rates shown are unscreened for the purpose of discussion. This heterostructure corresponds to $n_{2D}$ = 5$\times$10$^{12}$ cm$^{-2}$, $d$ = 3 nm, 20$\%$ Al in the alloy and roughness parameters: $L$ = 5 nm, $\delta$ = 0.5 nm.}
\end{figure*}

Fig.\ref{fig5}(a-c) show the 2D-2D scattering rates for different elastic scattering mechanisms for the first 5 subbands in the 2DEG at 300 K. The scattering rates shown are unscreened for the purpose of discussion in order to clearly understand the trend but a Thomas-Fermi type screening (only for intra-subband transitions as the inter-subband transitions involve large $q$) is used for the transport calculation. As discussed in the previous section (fig.\ref{fig2}(b)), as we move higher in subbands, the electrons become more delocalized as they move away from the interface and start behaving more like free electrons and hence increasing the probability in the bulk. At the same time, the probability near the interface decreases to keep the overall probability 1. This is being reflected in the scattering rates in the figure. Since the amount of wavefunction interacting with the impurity decreases near the interface, we see a lowering in the remote impurity scattering rate at higher subbands near the minima. The strong inverse $q$ dependence weakens the jumps due to density of states. Similarly, the interface roughness gives higher scattering rate initially at higher bands but then drops rapidly due to inverse $q$ dependence. A large drop at the minima for higher bands is due the effective electric field dependence which is maximum near the interface. Small jumps due to the density of states can be observed when looked carefully. Since, the electrons are more free to leak into the bulk than in the alloy (barrier), the total contribution (leakage) into the alloy decreases as we move higher in bands and hence the alloy disorder scattering rate is smaller for higher bands. Since, alloy disorder is isotropic in nature, we can clearly see jumps due to sudden changes in the density of states.

\subsubsection{Inelastic scattering rates: Polar optical phonon, Non-polar optical phonon $\&$ Acoustic phonon (2D-2D, 2D-3D, 3D-2D $\&$ 3D-3D)}

As stated before, POP modes limit the low field electron mobility in the bulk of $\beta$-G\MakeLowercase a$_{2}$O$_{3}$ and hence are expected to do the same in heterostructures. However, as already discussed, a very high electron density can be achieved in the 2DEGs of heterostructures, making the plasmon energy match with the energies of LO phonon modes. The two modes start influencing each other under resonance and form a set of coupled modes called LO phonon-plasmon coupled (LOPC) modes \cite{ref171,ref172,ref173}. This has been studied in several materials \cite{ref174,ref175,ref176,ref177,ref178,ref179,ref180,ref181,ref182} including the bulk of $\beta$-G\MakeLowercase a$_{2}$O$_{3}$ \cite{ref5,ref183,ref184}. At the same time, plasmon oscillating at comparable energy as LO phonons can screen ($\omega_{P}>\omega_{LO}$) the LO-TO splitting or enhances the LO phonon potentials strength through antiscreening ($\omega_{P}<\omega_{LO}$) \cite{ref173}. The dynamic screening in $\beta$-G\MakeLowercase a$_{2}$O$_{3}$ comes from of 12 IR active modes and high energy plasmon. 

The first order approximation for the 2D plasmons yield:
\begin{equation}\label{eq4.4}
    {\omega_{P}=\sqrt{\frac{\hbar^2n_{2D}e^2q}{m^*\epsilon_{\infty}}}}
\end{equation}
The LOPC modes corresponding to each $\vec{q}$ is calculated under plasmon-pole approximation given by \cite{ref5,ref189,ref190}:
\begin{equation}\label{eq4.5}
    {\epsilon_{\omega}(\vec{q})=\epsilon_{\infty}\prod_{i=1}^{12}\frac{(\omega_{i}^{LO}(\vec{q}))^2-\omega^2}{(\omega_{i}^{TO})^2-\omega^2}-\frac{\epsilon_{\infty}\omega_{P}^2}{\omega^2}}
\end{equation}

The scattering is still through LO modes as plasmons would only offer momentum exchange between the electrons rather than providing any average momentum relaxation for the ensemble. This means we must find the LO mode contribution to each LOPC modes.

The modified Fr\"ohlich vertex gives the scattering strength for the electron-phonon interaction \cite{ref5,ref189}:

\begin{widetext}
\begin{equation}\label{eq4.9}
    {\Big|M_{LOPC}^{\nu,LOj}(\vec{q})\Big|^2=\frac{e^2}{2\Omega\epsilon_{0}}\Bigg[\frac{\omega_{\nu}^{LOPC}(\vec{q})}{q^2}\Bigg(\frac{1}{\epsilon_{\omega_{\nu}^{LOPC}}^{-LOj}(\vec{q})}-\frac{1}{\epsilon_{\omega_{\nu}^{LOPC}}^{+LOj}(\vec{q})}\Bigg)\Lambda_{\nu}^{LOj}(\vec{q})\Bigg]}
\end{equation}
\end{widetext}
A pair of dielectric constants ${\varepsilon_{\omega_{v}^{LOPC}}^{+LOj}(\vec{q})}$ and ${\varepsilon_{\omega_{v}^{LOPC}}^{-LOj}(\vec{q})}$ must be calculated for each LO mode corresponding to a given LOPC mode. Here, ${\varepsilon_{\omega_{v}^{LOPC}}^{+LOj}(\vec{q})}$ includes the full response of that LO mode, while ${\varepsilon_{\omega_{v}^{LOPC}}^{-LOj}(\vec{q})}$ includes the response of all other modes keeping that LO mode frozen. 

At very high electron density, plasmons cease to behave as a collective excitation in the electron-hole pair continuum (EHC) and would affect the coupling. This is taken into account by taking off the plasmon dispersion from the calculation according to the upper boundary of EHC, given by \cite{ref191}: $\omega_{+}(q) = \frac{\hbar^{2}k_{F}q}{m^{*}} + \frac{\hbar^{2}q^2}{2m^{*}}$, $k_{F}$ is the Fermi vector at zero temperature.

\begin{figure}
\includegraphics[width=0.5\textwidth]{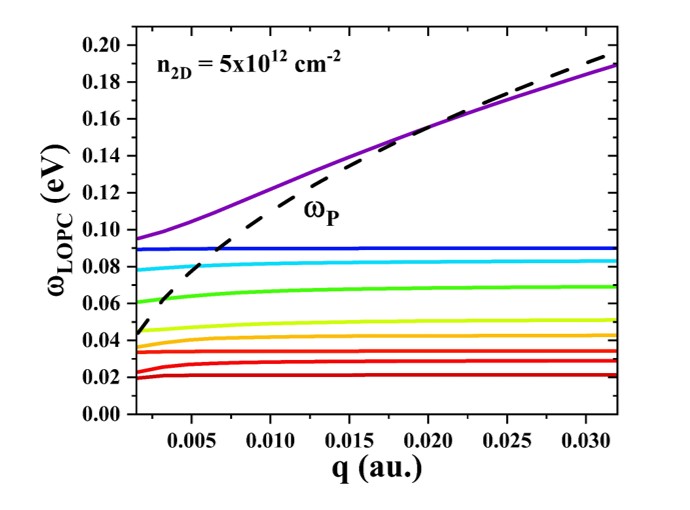}
\caption{\label{fig6} The LOPC frequencies calculated for n$_{2D}$=5$\times$10$^{12}$ cm$^{-2}$ corresponding to 8 LO modes polarized in x-z plane ($B_{u}$). The black dotted line represents the first order variation of plasmon frequency.}
\end{figure}

The scattering rate gets impacted by screening as well as the anti-screening of POP phonons through the 2D plasmons. Fig.\ref{fig6} shows the LOPC modes calculated for n$_{2D}$=5$\times$10$^{12}$ cm$^{-2}$ in cartesian x direction. There are 9 LOPC corresponding to 8 LO modes polarized in x-z plane ($B_{u}$). For very low electron density, when the plasmon energy is way below the range of POP energies, there is no screening involved and the corresponding scattering strength is unscreened. When the electron density is such that the corresponding plasmon energy is in the range of POP energies, the POP modes with lower energy get screened and the ones with higher energy get anti-screened. As the 2D plasmon energy is also proportional to the magnitude of the wavevector $\vec{q}$ to the first order of approximation, the number of modes getting screened increases with $q$. However, due to long range nature of such phonons, the scattering strength decreases and hence at low 2DEG densities, the anti-screening dominates and limits the overall low field mobility. However, at very high electron densities, the rate of increase in 2D plasmon energy is higher (higher slope) and hence even at smaller $q$, multiple modes get screened. 

\begin{figure*}
\includegraphics[width=\textwidth]{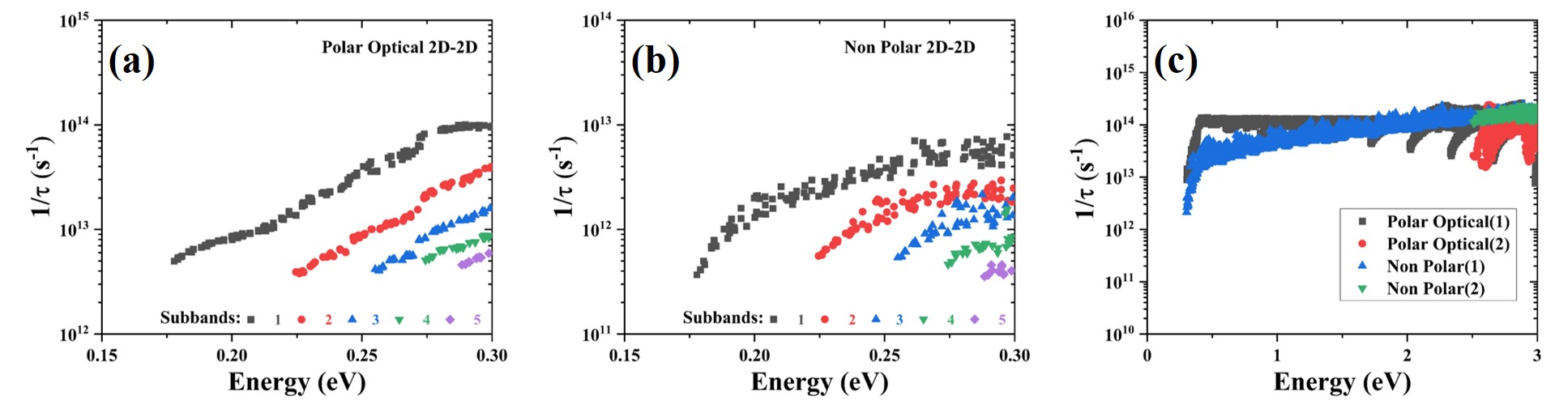}
\caption{\label{fig7} The 2D-2D (a) polar optical phonon (POP), and (b) non-polar (non-polar optical plus acoustic) phonon scattering rates (300 K) as a function of electron energy (eV) for the first 5 subbands. Anisotropy is clearly visible from the scattered points. The 0 eV is assumed to be the bottom of the 1st conduction band. This heterostructure corresponds to $n_{2D}$ = 5$\times$10$^{12}$ cm$^{-2}$, $d$ = 3 nm, 20$\%$ Al in the alloy and roughness parameters: $L$ = 5 nm, $\delta$ = 0.5 nm. (c) The 3D-3D polar optical phonon (Polar), and non-polar (non-polar optical plus acoustic) phonon scattering rates (300 K) as a function of electron energy (eV) for the first 2 conduction bands (denoted in brackets).  Anisotropy is clearly visible from the scattered points. The 0 eV is assumed to be the bottom of the 1st conduction band.}
\end{figure*}

Fig.\ref{fig7}(a-b) show the polar optical and non-polar phonon scattering rates (below 0.3 eV) for the first 5 subbands in the 2DEG at 300 K. A very usual trend can be seen due to their long and short range nature respectively. There is only absorption at lower energies which transforms to absorption and emission as the energy of the electrons keeps becoming greater than the energy of the phonon modes. Since, at a given energy, the number of available states in the same subband, (higher contribution due to higher overlap) after emitting a phonon, is more for a lower subband, the corresponding scattering rate is higher. Note that the final states fall only below 0.3 eV for 2D-2D scattering. The shown scattering rates correspond to $n_{2D}$ = 5$\times$10$^{12}$ cm$^{-2}$. At lower values of electron density, a higher POP scattering rate is expected and seen at higher energies as a result of antiscreening as discussed before. This would limit the electrons to drift into the bulk at higher fields which might impact the critical field value.

Fig.\ref{fig7}(c) shows the polar optical and non-polar phonon scattering rates (beyond 0.3 eV) for the first 2 conduction bands in bulk. The 3D scattering rates are only calculated until 3 eV following a previous Monte Carlo report for the bulk where the electron distribution drops to zero beyond 3 eV above electric field considered in this work \cite{ref18}. Note that the electrons near 0.3 eV can emit and jump to lower energies but that kind of transition is considered under 3D-2D scattering as discussed later. An increasing non-polar scattering rate with energy shapes the velocity-field curve at higher fields. The anisotropy in the scattering rates, coming from the anisotropy in the phonons, is clearly visible and is expected to impact our Monte Carlo calculations.

\begin{figure*}
\includegraphics[width=\textwidth]{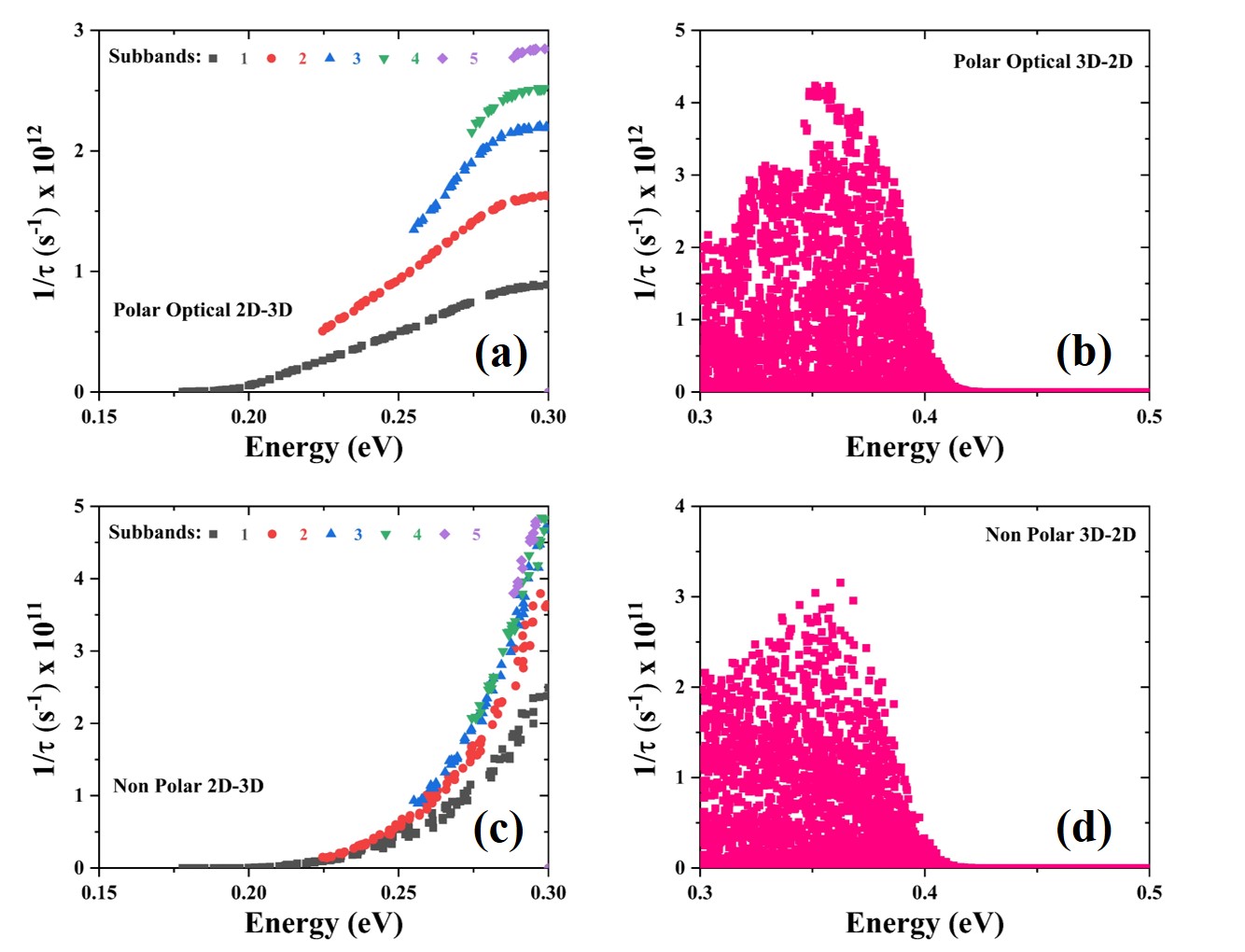}
\caption{\label{fig8} The (a) 2D-3D (300 K, only absorption) $\&$ (b) 3D-2D (300 K, only emission) polar optical phonon (POP), and (c) 2D-3D (300 K, only absorption) $\&$ (d) 3D-2D (300 K, only emission) non-polar (non-polar optical plus acoustic) phonon scattering rates as a function of electron energy (eV). Anisotropy is clearly visible from the scattered points. The 0 eV is assumed to be the bottom of the 1st conduction band. This heterostructure corresponds to $n_{2D}$ = 5$\times$10$^{12}$ cm$^{-2}$, $d$ = 3 nm, 20$\%$ Al in the alloy and roughness parameters: $L$ = 5 nm, $\delta$ = 0.5 nm.}
\end{figure*}

The escape and capture of electrons from and into the 2DEG is another important transition which must be considered. The 2D-3D and 3D-2D transitions happen though phonon absorption and emission respectively. This is usually ignored in simple Monte Carlo models for the 2DEG where such transitions are mediated only through drift \cite{ref201}. The scattering rate calculation follows the same process as 2D scattering with some modifications. The overlap integral is now between a confined wavefunction and a 3D wavefunction (more like a free electron). The 3D wavefunction for a given $k$ is calculated through Gram-Schmidt method \cite{ref207} such that it is orthogonal to each subband wavefunction. For 2D-3D scattering, the final state ranges from 0.3-0.5 eV (limited by maximum phonon energy) and is in the 3D k-space. For 3D-2D scattering, the initial 3D-wavevector again only ranges from 0.3-0.5 eV and the final state could lie in any of the subbands. The overlap integral takes care of the momentum conservation. Fig.\ref{fig8} shows the 2D-3D and 3D-2D scattering rates involving both polar optical (fig.\ref{fig8}(a-b)) and non-polar phonon (fig.\ref{fig8}(c-d)) modes. Since the electrons in the higher subbands are less confined and are more bulk like, the corresponding overlap between the 2D and the 3D wavefunction is large (better momentum conservation) increasing the scattering rate for the same \cite{ref202}. The 3D-2D transition shown correspond to each 3D k-point which would satisfy in-plane momentum conservation and have energies between (0.3-0.5) eV. The scattering rate decreases rapidly after $\sim$0.4 eV as the phonon emission (of max energy $\sim$0.1 eV) after this would make the final state fall in the bulk which is already considered in the 3D-3D scattering. A maximum near $\sim$0.35 eV is due to final states falling in the 5th subband when all the phonons can be emitted. It is interesting to note that due to bad momentum conservation the 3D-2D scattering rate is lower and hence would impact the carrier distribution by providing a small capture rate. 

\section{\label{sec:level2}Full-band Monte Carlo}

The full-band Monte Carlo (FBMC) approach \cite{ref208,ref209,ref210,ref211,ref212,ref213} is the most comprehensive and accurate method to solve the Boltzmann transport equation (BTE), governing the electron dynamics in the semi-classical regime. This method is an extension of general random sampling technique used to solve any multi-dimensional integral problem. For transport problems, a random walk is performed by the charge carriers to simulate their stochastic motion under the influence of several scattering processes. 

The idea is to simulate the carriers moving as free particles, subject to instantaneous random collisions. This is done through randomly generating free-flight times following \cite{ref211} $t_{0}=-\frac{1}{\Lambda}ln(r)$, where $r$ is a random number and $\Lambda$ is the maximum total scattering rate at any $k$ in a given subband/band. The random numbers, for a given core during a parallel computation, are computed by initially providing a random seed to avoid any duplication. The free particle motion is governed by \cite{ref211} $\vec{k_{new}}=\vec{k_{old}}-\frac{e\vec{E}t_{0}}{\hbar}$, where $k$ is the crystal momentum and $\vec{E}$ is the applied electric field in a given direction. If an electron in the 2DEG ends up at a higher energy than the cutoff during a drift process, a random number is used to select a given state such that the in-plane momentum and the total energy ($E_{tot}$) is conserved. Similarly, when an electron in the bulk loses energy during the drift process, the final subband $n$ is chosen such that the in-plane momentum is conserved and the total energy follows: $E_{n}<E_{tot}<E_{n+1}$. If the energy falls below that of the first subband, the electron is placed in the first subband. This, however, violates the energy conservation rule which is an assumption here. The drift process is followed by randomly choosing a scattering event through rejection technique and finally finding the final state with new momentum and energy. This is then repeated until a convergence in an observable is achieved. An ensemble of particles are simulated to study the time-dependent evolution of physically observable quantities such as average drift velocity, average energy, etc. These quantities are calculated at different sampled times where the motion of each particle is synchronized. 

An \textit{ab-initio} based FBMC simulator is developed from scratch, to investigate the high field transport in the 2DEG of heterostructures. The program is basically designed to perform three critical functions as discussed next:

\subsection{Initial distribution}

\begin{figure}
\includegraphics[width=0.5\textwidth]{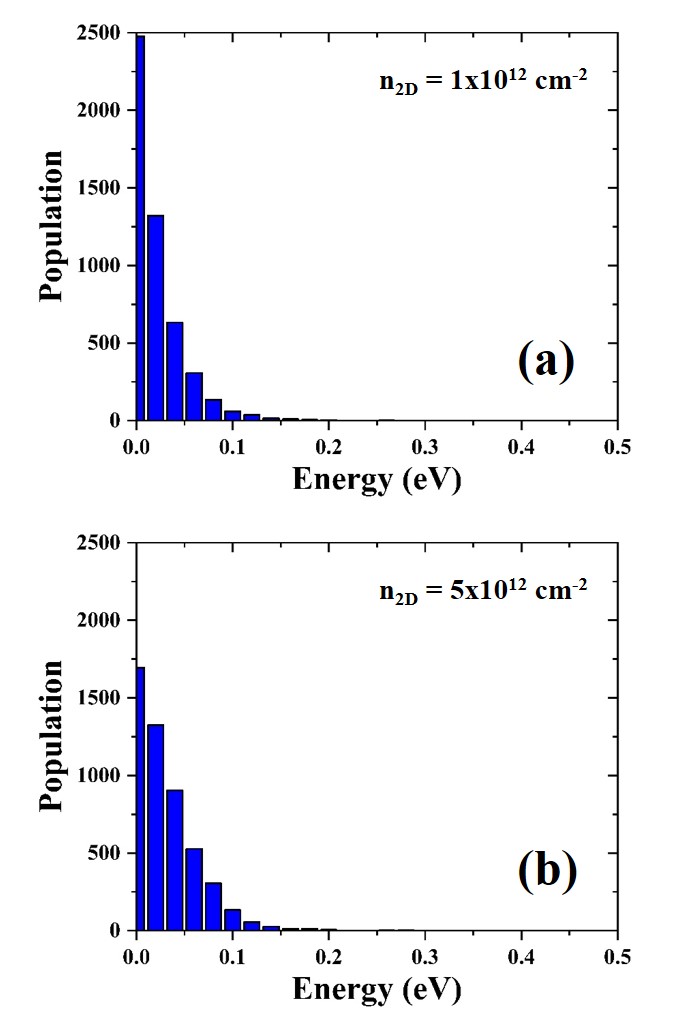}
\caption{\label{fig9} The initial distribution of 5000 electrons based on a Fermi-Dirac distribution function at 300 K for (a) $n_{2D}$ = 1$\times$10$^{12}$ cm$^{-2}$, and (b) $n_{2D}$ = 5$\times$10$^{12}$ cm$^{-2}$. The electrons are assumed to occupy only the first subband at zero electric field. The 0 eV is assumed to be the bottom of the first subband. Since the same number of electrons are simulated in both the cases, the electrons from the lower energy have moved to fill the higher energy states in the latter as the Fermi level moves up in the 1st subband.}
\end{figure}

The first step in the process is to initialize the particles based on Fermi-Dirac distribution. This approaches to Maxwell-Boltzmann distribution at high temperature and for non-degenerate doping. The Fermi-level is first calculated through ab-initio calculated density of states at a given temperature and electron density. A given number of particles are then distributed randomly such that they follow the same distribution. The wavector $\vec{k}$ corresponding to a given energy is then computed through a process similar to final state selection (explained later). Note that the electrons are assumed to occupy only the first subband when no field is applied in all cases. Fig.\ref{fig9} shows the distribution of 5000 electrons for $n_{2D}$ = 1$\times$10$^{12}$ cm$^{-2}$ and $n_{2D}$ = 5$\times$10$^{12}$ cm$^{-2}$ respectively scaled to satisfy the original distribution. As the latter is a degenerate case (Fermi level into the first subband), the probability of occupation is increased at higher energies \cite{ref15}. A fair comparison between the two would be to look at the change in distribution at different energy values rather than the actual population as the total number of carriers remain the same. After the electrons are initialized, the next steps are applied to individual electrons starting with the drift process through random flights, terminated by random scattering events. 
\subsection{Normalized scattering rates}
The next step is to select a scattering process from all mechanisms which would instantaneously scatter the particle and provide a new final state. The 2D and 3D scattering rates are calculated and stored for each $k$, initial and final subband/band and mode, type of mechanism (absorption/emission) in case of phonon involving scattering. The scattering rates corresponding to each mechanism are efficiently stored in a multidimensional array for each $k$ and subband/band and a separate tag keeps the information of the final state, mode and type of mechanism involved. Whenever a scattering mechanism has to be selected, a hashing method is used to quickly pickup all the scattering rates for a given $k$ and subband/band along with the tags. A given scattering is normalized as \cite{ref208}: 
\begin{equation}\label{eq5.1}
    S_{nrm}^{n}(m\vec{k}) = \frac{\sum_{i=1}^{n} S_{i}(m\vec{k})}{\Lambda}
\end{equation}
Where, $n$ is the nth mechanism, $m$ is the band/subband and $\Lambda$ is the sum of all possible scattering mechanisms at $m\vec{k}$. A given scattering mechanism is then stochastically selected using a random number such that $S_{nrm}^{n}(m\vec{k})>r>S_{nrm}^{n-1}(m\vec{k})$. The electron is considered to be self-scattered if $r>S_{nrm}^{p}(m\vec{k})$, where $p$ is the total number of scattering mechanisms, and the final state remains the same. 

The memory requirement is huge and increases with $N_{k}\times N_{p}\times N_{m}$ due to dense k-points, multiple bands and modes respectively but a heavy RAM for each core in a given node is utilized to cope with the same. 
\subsection{Final state selection}

The most computationally intensive step in the FBMC program is finding the final state of an electron once a scattering event has occured. Since, this involves many small sub-processes in order to accurately find the final state, the overall process is relatively slow and determines the total computation time of the FBMC program. Here, We discuss the elastic and inelastic scattering cases separately. 

\subsubsection{Elastic scattering}

The elastic scattering rate tag includes the information on final subband and hence the final state is searched through that particular band such that the momentum and energy is conserved. For isotropic scatterings, the angle between the initial and the final states is simply given by $\phi$=2$\pi$r, where, $r$ is a random number in (0,1) \cite{ref201}. For anisotropic scatterings, for a given $k$, the scattering rate is first calculated for a given set of discrete $q$ (determined by $\theta$) on a dense mesh and is then normalized by the total scattering rate as discussed before. A final state ($\theta$) is selected such that $S_{nrm}^{n}(m\vec{k})>r>S_{nrm}^{n-1}(m\vec{k})$, where $r$ is a random number in (0,1). This automatically takes into account the anisotropic nature of the scattering mechanism as each possible final states are weighted with the corresponding matrix element. The final state ($k_{x},k_{z}$) is then calculated by running an in-plane momentum conservation and total energy conservation on a fine grid. This final state is again limited by the density of chosen mesh. The next step would be to chose a random state within the square spanned (for a 2D k-space) by a given $\vec{k}$ point. This is explained later in this section. 

\subsubsection{Inelastic scattering}

\begin{figure*}
\includegraphics[width=0.9\textwidth]{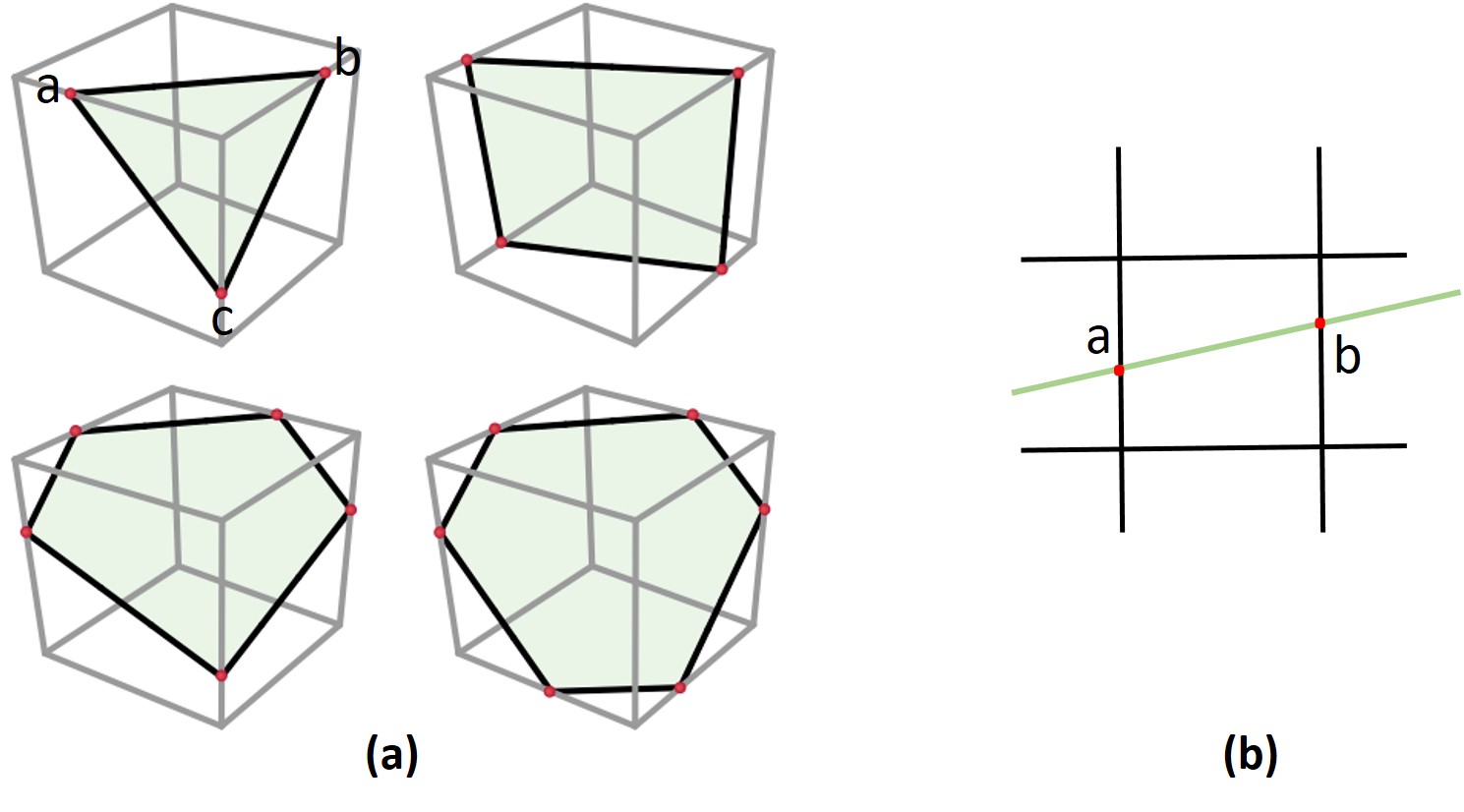}
\caption{\label{fig10} (a) The equienergy surfaces cutting a cube spanned by a k-point in the 3-D reciprocal space, with the contribution weight denoted by the shaded greenish region. (b) The equienergy line cutting a square spanned by a k-point in the 2-D reciprocal space, with the contribution weight denoted by the green line (between a-b). This is an important step in stochatically determining the final state which can be anywhere on the surface/line.}
\end{figure*}

Here, we explain the final state selection for a 3D region \cite{ref211} and a similar set of steps would apply in 2D but on a 2D k-space. First, all the k-points satisfying the energy and momentum conservation corresponding to a given mode ($\nu$) and mechanism (absorption+/emission-) within a certain smearing factor ($E_{mk_{i}}-E_{nk_{j}}\pm\hbar\omega_{\nu q}\leq0.01 eV$) are shortlisted. This is an intense process and requires heavy computing resources. A full phonon dispersion on a dense grid increases the complexity. An improvement can be achieved by hashing the band structure and phonon dispersion for an easy access. This gives all the $q$-points that would satisfy the energy-momentum conservation for a given mode ($\nu$). The next step is to find the matrix elements (EPI elements) $g(k_{i},k_{j}-k_{i})^{\nu}$ corresponding to the shortlisted $q$ points. This is done, as explained earlier, by storing the matrix elements which satisfy the energy-momentum conservation relation on an already defined grid. This significantly reduces the storage issue. Such matrix elements are stored in multiple files out of which a given file is first selected based on k-q hashing and then only the corresponding (to shortlisted points) file is read. Now, a final state ($k_{j}$) is picked stochastically, using a random number, based on the product of $|g(k_{i},k_{j}-k_{i})^{\nu}|^{2}$ and local density of states (LDOS) \cite{ref214}. Here, the LDOS is proportional to the area contributed by an equienergy surface ($E_{mk_{i}}\pm\hbar\omega_{\nu q}$) to a given cube (spanned by a k-point). This is shown in fig.\ref{fig10}(a). Once a cube is picked, the final state would lie inside that cube on that equienergy surface. Now, the shape of surface cut by the equienergy surface is determined followed by dividing it into smaller triangles. There are multiple shapes possible and hence the number of triangles could be more than 2. Now, A given triangle is selected stochastically using a random number. Once, we have a triangle as shown in fig.\ref{fig5.8}(a), the final state is given by \cite{ref215}: 
\begin{equation}\label{eq5.2}
    k_{f} = \vec{a} + \lambda_{1}(\vec{b}-\vec{a}) + \lambda_{2}(\vec{c}-\vec{a})
\end{equation}
Where, $\lambda_{1}=1-\sqrt{1-r}$, $\lambda_{2}=r(1-\lambda_{1})$ and $r$ is a random number in (0,1). 

The final state lies on a 2D k-space for the electrons in the 2DEG as shown in fig.\ref{fig10}(b). An equienergy line would cut the square spanned by a given k-point where the length of the line inside the square would correspond to the contributed LDOS. After selecting a single square, the final state is simply given by \cite{ref215}:
\begin{equation}\label{eq5.3}
    k_{f} = \vec{a} + \lambda_{1}(\vec{b}-\vec{a})
\end{equation}
When the electron is at higher energy and non-polar scattering rates become significant, the program starts accessing the matrix element files and the RAM requirement shoots up. This again increases the complexity of this step and the overall program run time. Several independent processes using sufficient RAM are run using mpi4py to better handle this issue.

\section{Results and Discussion}
The Monte Carlo simulation is performed with an ensemble of 5000 electrons for a certain number of cases involving different 2DEG densities and also an experimental structure for comparison and the corresponding results are presented here. We discuss the evolution of electron population and their transient dynamics with electric field for a given heterostructure with $n_{2D}$ = 1$\times$10$^{12}$ cm$^{-2}$ and $n_{2D}$ = 5$\times$10$^{12}$ cm$^{-2}$ respectively at 300 K. The electric field value ranges from 10 kVcm$^{-1}$ to 400 kVcm$^{-1}$ applied in cartesian -x direction. The two cases are compared to provide an insight on the effect of LOPC screening on POP scattering and hence on the behaviour of electron transport in the 2DEG. The heterostructure under consideration corresponds to $d$ = 3 nm, 20$\%$ Al in the alloy and roughness parameters: $L$ = 5 nm, $\delta$ = 0.5 nm. However, the velocity field curves are discussed for a few more cases to provide a fair comparison between our calculation and the available experimental data.  
 
\subsection{Zone population}

\begin{figure*}
\includegraphics[width=\textwidth]{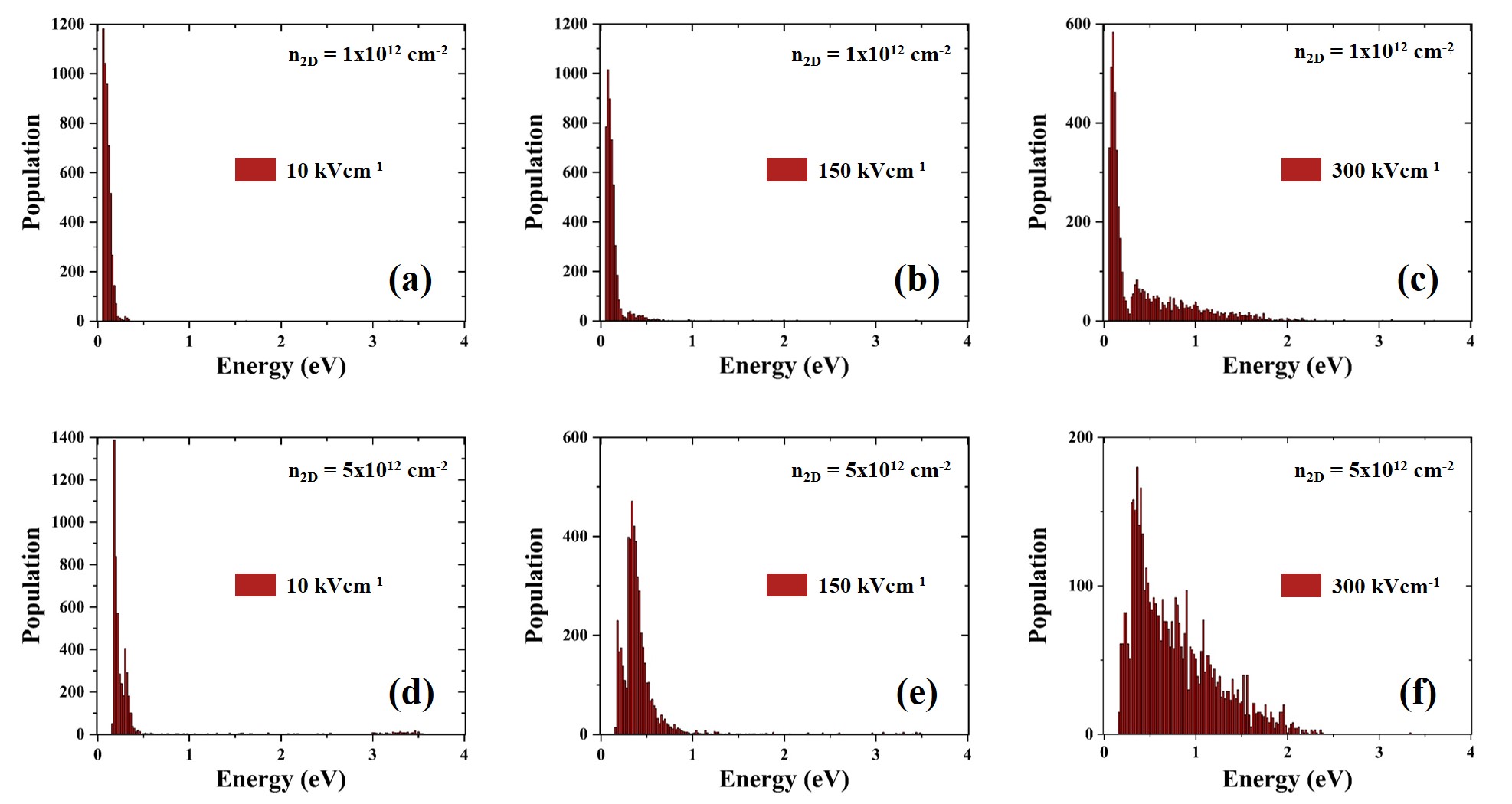}
\caption{\label{fig11} (a-c) $\&$ (d-f) show the steady state electron population with energy (eV) at electric field of 10, 150 and 300 kVcm$^{-1}$ (in -x direction) for (a-c) $n_{2D}$ = 1$\times$10$^{12}$ cm$^{-2}$, and (d-f) $n_{2D}$ = 5$\times$10$^{12}$ cm$^{-2}$ respectively. The screening in the latter case causes more number of electrons to occupy the higher energy states at higher fields. A kink at the 2D-3D boundary is a result of low 3D-2D scattering rate making it difficult for the electrons to bounce back into the 2DEG.}
\end{figure*}

Under high field, the electrons have distribution covering the whole brillouin zone. Fig.\ref{fig11}(a-f) show the evolution of an ensemble of electrons with energy at electric fields: 10 kVcm$^{-1}$, 150 kVcm$^{-1}$ and 300 kVcm$^{-1}$ for electron densities: $n_{2D}$ = 1$\times$10$^{12}$ cm$^{-2}$ and $n_{2D}$ = 5$\times$10$^{12}$ cm$^{-2}$ respectively. The initial distribution function follows Fermi-Dirac at zero field as discussed before. The distribution is collected once the steady state has reached at a given applied electric field. The distribution dies off quickly after $\sim$2 eV in both the cases even at 300 kVcm$^{-1}$. The velocity saturation results from the short range intra-valley EPI. The satellite valley, as discussed before, lies around $\sim$2.5 eV and hence the intervalley scattering is not responsible for negative differential conductance (NDC). The non-parabolicity of the conduction band at higher energies reduces the average electronic velocities and results in the NDC unlike the mechanism observed in the bulk of GaAs and GaN (due to intervalley scattering).

The antiscreening offered by LOPC phonon in the former case ($n_{2D}$ = 1$\times$10$^{12}$ cm$^{-2}$) at higher energies, as discussed before, causes high 2D-2D scattering in the 2DEG and hence decreases the drift rate. Comparatively, the screening dominates when $n_{2D}$ = 5$\times$10$^{12}$ cm$^{-2}$ and hence the drift rate is larger shaping the distribution function accordingly. A kink near 0.3 eV, as seen in both the cases, is due to low 3D-2D scattering rate causing it difficult for the electrons to get captured back into the 2DEG. 

\begin{figure*}
\includegraphics[width=\textwidth]{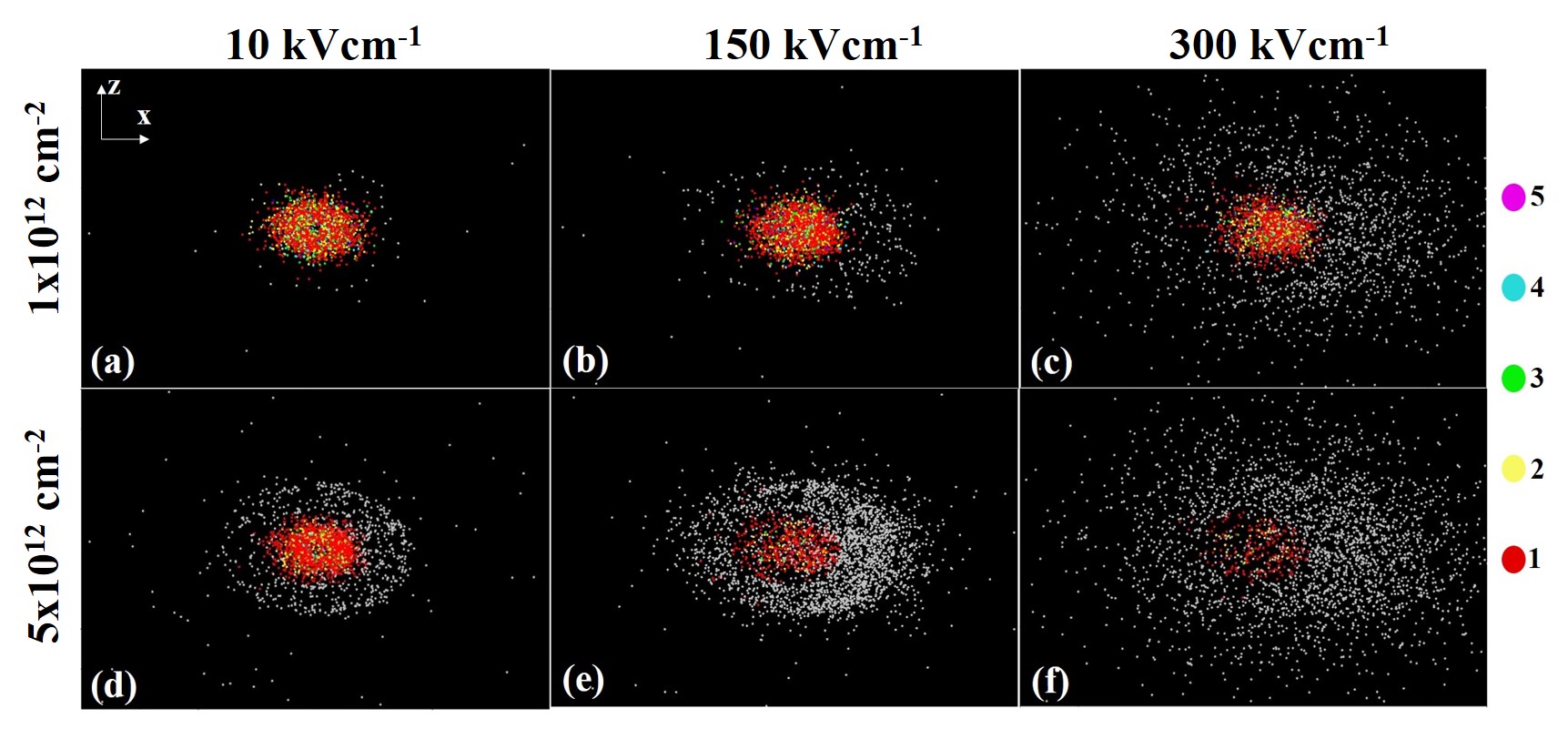}
\caption{\label{fig12} (a-c) show the evolution of electron population (steady state) in all the 5 subbands (colored) and in the bulk (gray) in a 2D k-space for $n_{2D}$ = 1$\times$10$^{12}$ cm$^{-2}$ at electric field of (a) 10, (b) 150 and (c) 300 kVcm$^{-1}$ (in -x direction). (d-f) show the evolution of electron population (steady state) in all the 5 subbands (colored) and in the bulk (gray) in a 2D k-space for $n_{2D}$ = 5$\times$10$^{12}$ cm$^{-2}$ at electric field of (d) 10, (e) 150 and (f) 300 kVcm$^{-1}$ (in -x direction). Note that electrons in the bulk have all three momentum components but the $k_{y}$ is omitted in the figure for presentation. The overall shift in +x direction due to field (in -x) can be clearly seen. Also, the population in the 2DEG (colored dots correspond to different subbands: see legend) is seen to be decreasing with the field as electrons become more bulk type.}
\end{figure*}

The distribution in k-space once the steady state has reached is shown in fig.\ref{fig12}(a-c)$\&$(d-f) when the field is applied in -x direction for $n_{2D}$ = 1$\times$10$^{12}$ cm$^{-2}$ and $n_{2D}$ = 5$\times$10$^{12}$ cm$^{-2}$ respectively. The colored dots correspond to the electrons in the 2DEG (x-z component) whereas the gray dots are the 3-D electrons with $y$ momentum component omitted for a better presentation. An almost symmetric distribution at 10 kVcm$^{-1}$ shifts to x region when higher field is applied. A significant population in the -x direction is due to high non-polar momentum relaxation rate as compared to low energy relaxation. The net current is due to the contribution from the overall distribution. Another interesting thing to notice here is shift in the electron population from the 2DEG subbands  to the bulk with applied electric field. The shown region only covers 40$\%$ of the entire brillouin zone and hence it can be inferred that the zone edges are not populated with enough electrons due to the high non-polar scattering rates.

\subsection{Transient dynamics}
The transient dynamics in Monte Carlo simulation becomes very important when studying a scaled device where the channel length is short enough for the electrons to not reach the steady state while drifting from source to the drain terminal. For applications in high power electronics, this won't play a significant role. However, as we have thoroughly discussed before that $\mathrm{\beta-Ga_{2}O_{3}}$ finds its application in RF switching as well, where the scaling of the devices matters, hence making it critical to discuss such characteristics here. 

\begin{figure}
\includegraphics[width=0.5\textwidth]{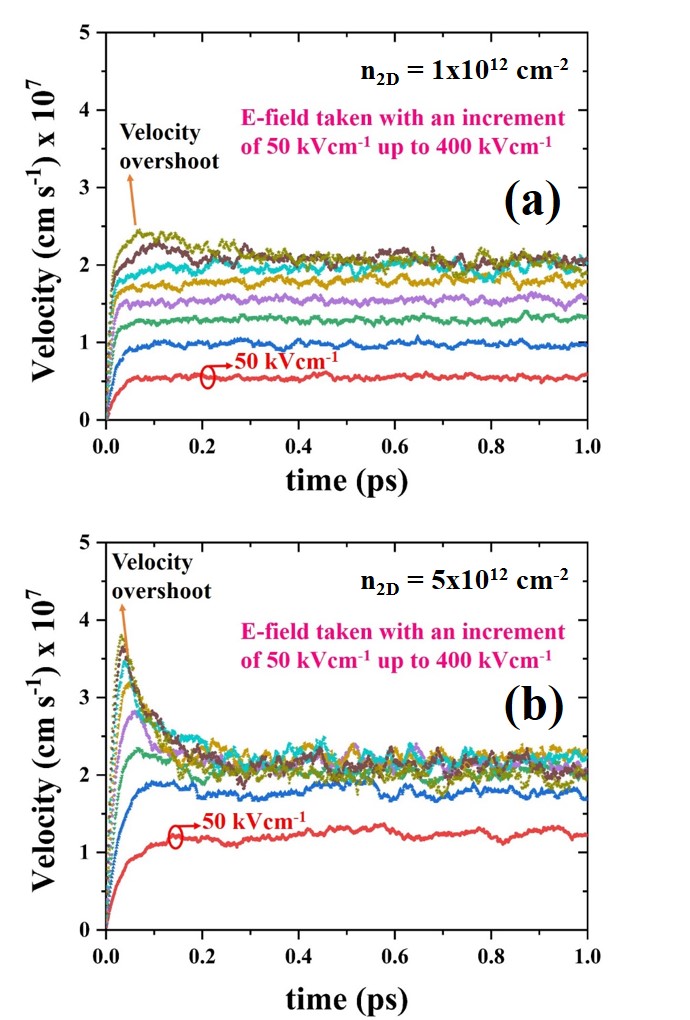}
\caption{\label{fig13} The transient characteristics of an ensemble of 5000 electrons under an influence of a range of electric field values for (a) $n_{2D}$ = 1$\times$10$^{12}$ cm$^{-2}$, and (b) $n_{2D}$ = 5$\times$10$^{12}$ cm$^{-2}$. The electric field varies as 50:50:400 kVcm$^{-1}$. The observed NDC with the field is due to the non-parabolicity of $\Gamma$ valley at higher energies.}
\end{figure}

The velocity-time plot for $n_{2D}$ = 1$\times$10$^{12}$ cm$^{-2}$ and $n_{2D}$ = 5$\times$10$^{12}$ cm$^{-2}$ at different electric field values (10,50,100,150,200,250,300,350,400) kVcm$^{-1}$ is shown in fig.\ref{fig13}(a)$\&$(b) respectively. An ensemble of 5000 electrons is simulated and the velocity of each of them is extracted and then averaged out when their motion is synchronized (at a given time). At lower electric fields, the POP scattering is dominant and hence a smooth transition to a steady state can be seen in the velocity. This is due to low momentum and energy relaxation provided by polar optical phonons. At higher fields, when the electrons are in the bulk, the non-polar scattering starts kicking in and rapidly increases the momentum relaxation rate while keeping the energy relaxation still limited by the available phonon energies. This causes the velocity to shoot up due to lower momentum randomization (limited by POP) and then drop down to attain a steady state due to higher momentum randomization (limited by non-polar phonons). On the other hand, there is a smooth change in the energy until the steady has reached. The velocity overshoot in the former case (a) is not seen until a high field (300 kVcm$^{-1}$) is applied. This is due to high POP scattering rate (antiscreening from LOPC) in the 2DEG. However, the more screening in the latter case (b) enables the electrons to enter into the bulk at even lower field (150 kVcm$^{-1}$) and get impacted by the non-polar scattering, causing an early velocity overshoot. A device length defined by the area under the overshoot curve can take advantage of the high velocity for RF switching applications. This, however, would require higher 2DEG density and smaller device length for the screening to dominate and provide higher velocity as seen in the figure.  

\subsection{Velocity-field curves}
\begin{figure}
\includegraphics[width=0.5\textwidth]{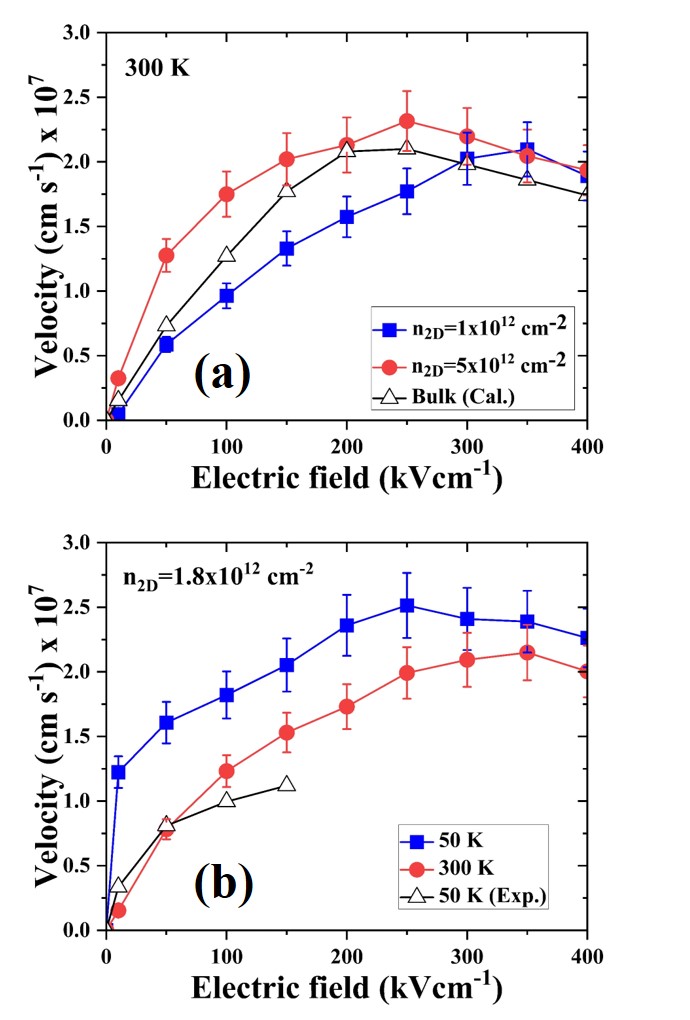}
\caption{\label{fig14} (a-b) Velocity-field curves for a few cases (TW: This Work) as seen in the legend of the figure. A comparison with the experiment (Exp.) \cite{ref17} and calculated (Cal) bulk values \cite{ref18} is also shown. The spacer thickness in heterostructrues in (b) is increased from 3 nm to 4.5 nm for a fair comparison with the experiment. The error bars correspond to 10$\%$ variation in the electron effective mass.}
\end{figure}

Fig.\ref{fig14} shows a comparison between velocity-field curves for a few heterostructures and the bulk $\mathrm{\beta-Ga_{2}O_{3}}$. This is an important topic of discussion as the characteristics curves of a device depend on the impact of high field on the velocity of carriers. The experimental heterostructure is similar to the one used in this work with spacer thickness changed to $d$ = 4.5 nm to give an electron density of $n_{2D}$ = 1.8$\times$10$^{12}$ cm$^{-2}$ in the 2DEG \cite{ref17}. Before further discussion, it's important to reiterate here that the velocity saturation is resulting from intravalley non-polar scattering in the bulk region and the NDC seen is simply a reflection of non-parabolicity of the $\Gamma$ valley at higher energies. The net velocity is the contribution from all the electrons and hence is a strong function of distribution in the momentum-energy space.

As seen in fig.\ref{fig14}(a), the peak corresponding to $n_{2D}$ = 5$\times$10$^{12}$ (calc.) lies at around the same field (250 kVcm$^{-1}$) as bulk but with higher velocity ($\sim$2.25$\times$10$^{6}$ cms$^{-1}$). This is due to screening present in the 2DEG reducing the corresponding population at lower energies. The corresponding low field (10 kVcm$^{-1}$) mobility is ($\sim$393 cm$^{2}$V$^{-1}$s$^{-1}$). The velocity at the same field is low for $n_{2D}$ = 1$\times$10$^{12}$ (calc.) at 300 K and peaks at 350 kVcm$^{-1}$. This is due to the higher electron population in the 2DEG (low energies) even at 250 kVcm$^{-1}$ as seen before (due to higher POP scattering) making the overall velocity low. The peak velocity is comparable to bulk for $n_{2D}$ = 1$\times$10$^{12}$ and slightly higher for $n_{2D}$ = 1.8$\times$10$^{12}$ due to their respective low field (10 kVcm$^{-1}$) mobility values of ($\sim$107 cm$^{2}$V$^{-1}$s$^{-1}$) and ($\sim$190 cm$^{2}$V$^{-1}$s$^{-1}$) respectively.

The low field (10 kVcm$^{-1}$) mobility corresponding to the calculated experimental structure at 50 K is $\sim$1220 cm$^{2}$V$^{-1}$s$^{-1}$, and the velocity peaks at 250 kVcm$^{-1}$ around ($\sim$2.5$\times$10$^{6}$ cms$^{-1}$) coming from higher low field mobility. This is shown in fig.\ref{fig5.14}(b). The discrepancy between the two could be attributed to the contact resistance contribution in the measured values which would compress the overall plot along the field axis. The other possible reason could be the self-heating effect [ref], which is ignored in this work and is supposed to decrease the net velocity at higher fields. The error bars correspond to 10$\%$ variation in the electron effective mass.  

\section{Conclusion}

The high field electron transport in the 2DEG of $\beta$-(A\MakeLowercase l$_{\MakeLowercase x}$G\MakeLowercase a$_{1-\MakeLowercase x}$)$_{2}$O$_{3}$/G\MakeLowercase a$_{2}$O$_{3}$ heterostructures is investigated using full-band Monte Carlo approach. An in-house developed program is utilized to extract the parameters under interest such as velocity field curves, velocity-time plots etc which can ultimately be used to design improved devices for better performance. A comparison between a few heterostructure devices and the bulk is presented with the maximum velocity reaching up to $\sim$2.25$\times$10$^{6}$ cms$^{-1}$ at 300 K for $n_{2D}$ = 5$\times$10$^{12}$ with the electric field value of 250 kVcm$^{-1}$ comparable to bulk.

\begin{acknowledgments}
The authors acknowledge the support from Air Force Office of Scientific Research under award number FA9550-18-1-0479 (Program Manager: Ali Sayir) and from NSF under award ECCS-2019749, from Semiconductor Research Corporation under GRC
Task ID 3007.001. The authors also acknowledge the high performance computing facility provided by the Center for Computational Research (CCR) at University at Buffalo.
\end{acknowledgments}

\section*{Data Availability Statement}
The data and the in-house developed programs that support the findings of this study are available from the corresponding author upon reasonable request. The \textit{ab-initio} calculations are performed using the open source software, Quantum Espresso. The licensed version of Silvaco Atlas is used for the self-consistent Schoringer-Poisson calculations.

\nocite{*}
\bibliography{aipsamp}

\end{document}